\documentclass[final,1p,times]{elsarticle} 
\usepackage{graphicx} 
\usepackage{amssymb} 
\usepackage{amsthm} 
\usepackage{lineno} 

\journal{Nuclear Physics A} 
\begin{document} 

\begin{frontmatter} 
%
\vspace{12pt}



\title{Chemical Equilibration and Transport Properties of Hadronic Matter near $T_c$}

\author{ J.~Noronha-Hostler$^{a}$$^{b}$$^{c}$, J.~Noronha$^{c}$, H.~Ahmad$^{a}$, I.~Shovkovy$^{d}$, C.~Greiner$^{a}$}

\address[a]{Institut f\"ur Theoretische Physik, J. W.
Goethe Universit\"at, D-60438 Frankfurt am Main, Germany }

\address[b]{Frankfurt Institute for Advanced Studies, J. W.
Goethe Universit\"at, D-60438 Frankfurt am Main, Germany}

\address[c]{Department of Physics, Columbia University, New York, NY 10027 USA}

\address[d]{Arizona State University, Mesa, AZ 85212 USA}


\begin{abstract} 

We discuss how the inclusion of Hagedorn states near $T_c$ leads to short chemical equilibration times of proton anti-proton pairs, $K\bar{K}$ pairs, and $\Lambda\bar{\Lambda}$ pairs, which indicates that hadrons do not need to be ``born" into chemical equilibrium in ultrarelativistic heavy ion collisions. We show that the hadron ratios computed within our model match the experimental results at RHIC very well. Furthermore, estimates for $\eta/s$ near $T_c$ computed within our resonance gas model are comparable to the string theory viscosity bound $\eta/s=1/4\pi$. Our model provides a good description of the recent lattice results for the trace anomaly close to $T_c=196$ MeV. 
\end{abstract}

\end{frontmatter} 


\section{Introduction}
Hagedorn states were originally introduced to explain the exponentially increasing mass spectrum in $p-p$ and $p-\pi$ scatterings. Hagedorn fit the mass spectrum up to $\Delta (1232)$ within the statistical bootstrap model \cite{Hagedorn:1968jf} in the late 60's and an update on the experimental verification of this asymptotic behavior can be found in \cite{Broniowski:2004yh}. In Hagedorn's picture of a hadron gas, as the energy is increased the number of degrees of freedom increases, not the average momentum per particle. In this case, such rapidly increasing density of states leads to a ``limiting" (Hagedorn) temperature, $T_{H}$, beyond which ordinary hadronic matter cannot exist \cite{Hagedorn:1968jf}. 

The presence of Hagedorn states opens up the phase space for multiple particle decays \cite{NoronhaHostler}. These multiple particle collisions can in turn produce an $X\bar{X}$ through the following decay
\begin{equation}\label{eqn:reaction}
n\pi\leftrightarrow HS\leftrightarrow n^{\prime}\pi+X\bar{X}
\end{equation}   
where $X\bar{X}=p\bar{p}$, $K\bar{K}$, or $\Lambda\bar{\Lambda}$. The motivation behind Eq.\ \ref{eqn:reaction} stems from the inability to explain particle yield ratios at RHIC using known decay channels of both binary and multiple mesonic reactions. Initially, it was suggested that the hadrons must then be ''born" in chemical equilibrium because the time scales were too long \cite{bornequilibrium}. However, in this paper we discuss how Hagedorn states are able to populate  $X\bar{X}$ pairs before chemical freezeout.  We use rate equations to describe the behavior of the Hagedorn states within the hadron phase. Additionally, our model including Hagedorn states provides a good description \cite{NoronhaHostler:2008ju} of the lattice results for the thermodynamic properties \cite{Cheng:2007jq,zodor} of the quark-gluon plasma (QGP). These highly massive states should also contribute to the transport properties of the QGP and, in fact, here we will show that the inclusion of Hagedorn states near $T_c$ can lower $\eta/s$ to be close to the string theory bound $\eta/s=1/4\pi$ \cite{Kovtun:2004de}. Furthermore, Hagedorn states provide a unique method to distinguish between different critical temperatures in the context of thermal fits to particle yields \cite{NoronhaHostler:2009tz}.  
 
\section{Model}\label{sec:model}

Hagedorn states in our extended Hagedorn gas model are included in our model via the exponentially rising mass spectrum of hadrons, 
\begin{equation}\label{eqn:fitrho}
    \rho=\int_{M_{0}}^{M}\frac{A}{\left[m^2 +m_{r}^2\right]^{\frac{5}{4}}}e^{\frac{m}{T_{H}}}dm.
\end{equation}
where $M_{0}=2$ GeV and $m_{r}^2=0.5$ GeV. We consider here only mesonic, non-strange Hagedorn states. We assume that $T_H=T_c$, and then we consider the two different different lattice results for $T_c$: $T_c=196$ MeV \cite{Cheng:2007jq} and $T_c=176$ MeV \cite{zodor}. Furthermore, repulsive interactions among the hadrons are taken into account via the excluded-volume corrections worked out in \cite{kapustavolcorr}. 

In order to find the maximum Hagedorn state masses $M$  and the ''degeneracy" A, we fit our model to the thermodynamic properties of the lattice \cite{Cheng:2007jq,zodor}. In \cite{Cheng:2007jq} the thermodynamical properties are derived from the trace anomaly $\varepsilon-3p$, which is what we fit in order to obtain the parameters for the Hagedorn states. Thus, we set $T_c=196$ MeV, $A=0.5 \,{\rm GeV}^{3/2}$, $M=12$ GeV, and $B=\left(340 \,{\rm MeV}\right)^4$, which plays the role of an effective bag constant. The fit for the trace anomaly $\Theta/T^4$ is shown in \cite{NoronhaHostler:2008ju}. Our curves provide a good description of the lattice results. For the lower critical temperature found Ref.\ \cite{zodor} we fit the energy density, as is shown in \cite{longpaper}. From that we obtain $T_c=176$ MeV, $A=0.1 \,{\rm GeV}^{3/2}$, $M=12$ GeV, and $B=\left(300\, {\rm MeV}\right)^4$, which provides a good fit for the lattice data.

\section{Results}

Using our Hagedorn gas model \cite{NoronhaHostler:2009tz}, we first calculated the thermal fits for a hadron gas without Hagedorn states, such that $T_{ch}=160.4$ MeV, $\mu_b=22.9$ MeV, and $\chi^2=21.2$, which matches extremely well the results of Ref.\ \cite{tf}. 

\begin{figure*}[t]
\begin{minipage}[b]{0.45\linewidth}
\centering
\includegraphics[width=1\textwidth]{tfHS.eps}
\caption{Thermal fits for the to the experimental data \cite{STAR,PHENIX} with the inclusion of Hagedorn states.} \label{fig:tfHS}
\end{minipage}
\hspace{0.5cm}
\begin{minipage}[b]{0.45\linewidth}
\includegraphics[width=1\textwidth]{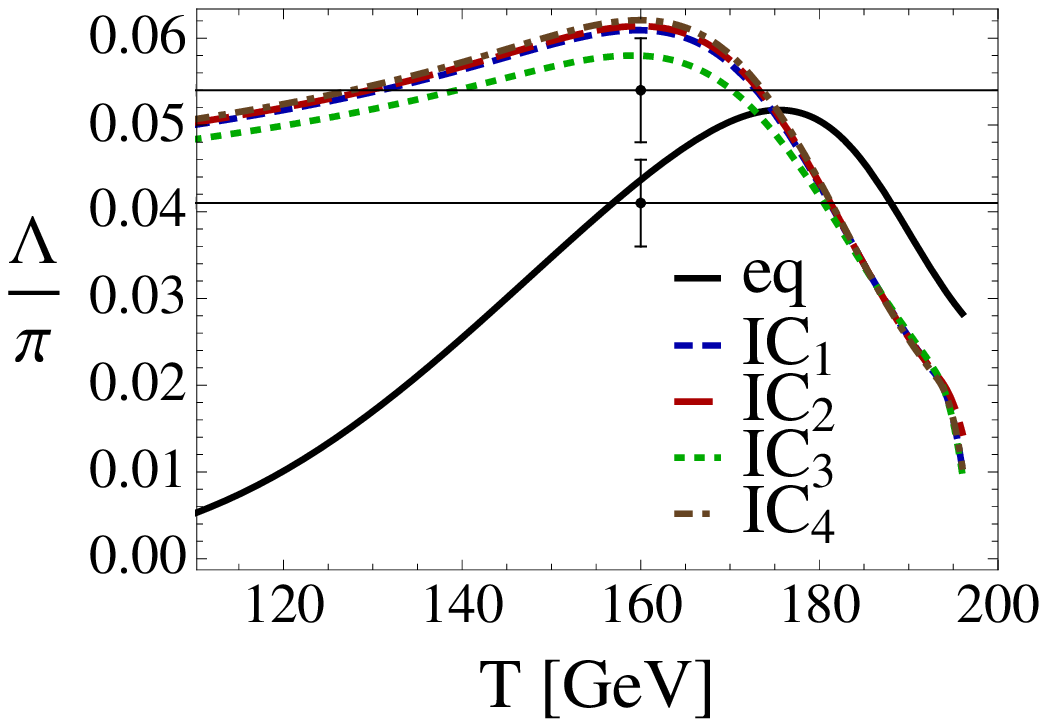}
\caption{Ratio of $\Lambda/\pi$ for various initial conditions when $T_H=196$ MeV compared to the experimental data \cite{STAR}.} \label{fig:ll}
\end{minipage}
\end{figure*}

With the inclusion of Hagedorn states (see \cite{NoronhaHostler:2009tz}) the fit for the RBC-Bielefeld collaboration is $T_{ch}=165.9$ MeV, $\mu_b=25.3$ MeV, and $\chi^2=20.9$, which is shown in Fig.\ \ref{fig:tfHS}.  When we consider the lattice results from \cite{zodor}, which are at the lower end of the critical temperature spectrum ($T_c=176$ MeV), we find $T_{ch}=170.4$ MeV, $\mu_b=27.8$ MeV, and $\chi^2=18.8$. The lower critical temperature seems to have a significant impact on the thermal fit. The lower $\chi^2$ is due to the larger contribution of Hagedorn states at $T_{ch}=170.4$ MeV, which is much closer to the $T_c$ found in \cite{zodor}.  

\begin{figure}
\centering
\includegraphics[width=0.45\textwidth]{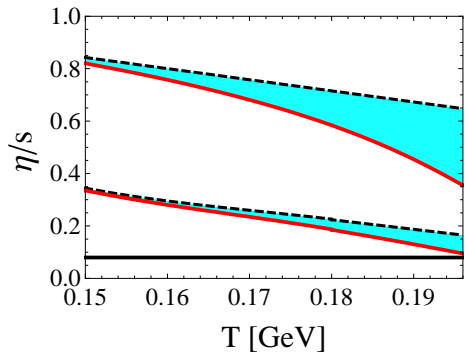}
\caption{$\eta/s$ is shown for a gas of pions and nucleons \cite{Itakura:2007mx} (upper dashed black line) and for a hadron resonance gas with (constant) excluded volume corrections \cite{Gorenstein:2007mw} (lower dashed black line). An upper bound on the effects of HS on $\eta/s$ is shown in solid red lines. The blue band between the curves is used to emphasize the effects of HS. The solid black line at the bottom is the AdS/CFT lower bound $\eta/s=1/4\pi$ \cite{Kovtun:2004de}.} \label{fig:etas}
\end{figure}

When we use Eq.\ \ref{eqn:reaction} to dictate the dynamics (see \cite{NoronhaHostler,longpaper}) of chemical equilibration, we find that the $X\bar{X}$ pairs reach chemical equilibrium before $T=160$ MeV. As shown in Fig.\ \ref{fig:ll}, the $\Lambda\bar{\Lambda}$ pairs reach the experimental values already at about $T=180$ MeV. In \cite{longpaper} a similar analysis is shown for the $p\bar{p}$ and $K\bar{K}$ pairs when $T_c=196$ MeV and $T_c=176$. Moreover, the contribution of the Hagedorn states to the upper bound on $\eta/s$ can be calculated using
\begin{eqnarray}\label{etas1}
\left(\frac{\eta}{s}\right)_{tot}&\leq&\frac{s_{HG}}{s_{HG}+s_{HS}}\left[\left(\frac{\eta}{s}\right)_{HG}+ \frac{\eta_{HS}}{s_{HG}}\right].
\end{eqnarray}
as derived in \cite{NoronhaHostler:2008ju}. The results are shown in Fig.\ \ref{fig:etas}, which reveals the drop of $\eta/s$ to the string theory limit \cite{Kovtun:2004de}. A discussion of the effects of Hagedorn states on the bulk viscosity can be found in \cite{NoronhaHostler:2008ju}.

\section{Conclusions}

In conclusion, in this paper we showed that Hagedorn states may play a fundamental roll in the dynamics of the QGP close to $T_c$.  They allow for short chemical equilibrium times so that the hadrons can be ''born" out of chemical equilibrium and dynamically reach chemical equilibrium within the hadronic fireball. The $\Lambda/\pi$ ratio was shown to reproduce the experimental data \cite{STAR}.  Additionally, Hagedorn states improve the fit to particle yield ratios in thermal models and also slightly increase the chemical freeze-out temperature.  These two results are consistent with each other because the dynamics of Hagedorn states lead to such quick chemical equilibrium times that the chemical freeze-out temperature is higher than in the scenario when no Hagedorn states are included.  Additionally, the $\chi^2$ for the thermal fits depends on $T_c$ and lower $\chi^2$ values are found for the lower $T_c$ considered here. Using the fit to the lattice data for $T_c=196$ MeV we also find that $\eta/s$ can be comparable to the string theory bound already in the hadronic phase near $T_c$.  Clearly, our results indicate that Hagedorn states may play a significant role in the dynamics of the QGP close to $T_c$ and must be taken into account in future studies of the hadronic properties of the quark-gluon plasma formed in heavy ion collisions.

\section{Achknowledgements}

J.N.\ acknowledges support from US-DOE Nuclear Science Grant No.\
DE-FG02-93ER40764. This work was supported by the Helmholtz International Center for FAIR within the LOEWE program launched by the State of Hesse.

\end{document}